\documentclass[aps,prb,reprint]{revtex4-1}
\usepackage{color}
\usepackage{graphicx}
\usepackage{hyperref}

\begin{document}

\title{Synthesis of nanocrystalline titanium nitride coatings from the plasma of a composite-cathode arc discharge}

\author{Yu.~F.~Ivanov}
 \affiliation{Institute of High Current Electronics, SB RAS, 2/3 Akademichesky Avenue, Tomsk, 634055, Russia}
\author{N.~N.~Koval}
 \affiliation{Institute of High Current Electronics, SB RAS, 2/3 Akademichesky Avenue, Tomsk, 634055, Russia}
\author{O.~V.~Krysina}
 \affiliation{Institute of High Current Electronics, SB RAS, 2/3 Akademichesky Avenue, Tomsk, 634055, Russia}
\author{T.~Baumbach}
 \affiliation{Karlsruhe Institute of Technology (KIT), Institute for synchrotron radiation, Hermann-von-Helmholtz-Platz 1, 76344 Eggenstein-Leopoldshafen, Germany, EU}
\author{ S.~Doyle}
 \affiliation{Karlsruhe Institute of Technology (KIT), Institute for synchrotron radiation, Hermann-von-Helmholtz-Platz 1, 76344 Eggenstein-Leopoldshafen, Germany, EU}
\author{T.~Slobodskyy}
 \email{Taras.Slobodskyy@kit.edu}
 \affiliation{Karlsruhe Institute of Technology (KIT), Institute for synchrotron radiation, Hermann-von-Helmholtz-Platz 1, 76344 Eggenstein-Leopoldshafen, Germany, EU}
\author{N.~A.~Timchenko}
 \affiliation{Tomsk Polytechnic University, Tomsk, 634050, Russia}
\author{R.~M.~Galimov}
 \affiliation{Tomsk Polytechnic University, Tomsk, 634050, Russia}
\author{I.~P.~Chernov}
 \affiliation{Tomsk Polytechnic University, Tomsk, 634050, Russia}
\author{ A.~N.~Shmakov}
 \affiliation{Siberian synchrotron radiation centre, Budker Institute of Nuclear Physics, SB RAS, Lavrentyev av. 11, Budker INP, Novosibirsk, 630090, Russia}

\begin{abstract}
Experimental data are given on the structure and properties of nanocrystalline hardening coatings of titanium nitride doped with copper, produced by plasma-assisted vacuum arc deposition by evaporating powder cathodes. A model of nanostructurization of this type of coatings is proposed. According to the model, the nanocrystallization in these materials is due to the dopant atoms, which form an amorphous sheath around a crystallite, thus defining its size.
\end{abstract}


%81.07.Bc - {Nanocrystalline materials}
%68.55.A- - {Nucleation and growth}
%68.37.Lp - {Transmission electron microscopy (TEM)}
%61.05.C- - {X-ray diffraction}
%78.70.En - {X-ray emission spectra and fluorescence}

%\keywords{synchrotron; X-ray diffraction; fluorescence; TEM; nanocrystal, growth;}

\maketitle

\section{Introduction}

The vacuum-arc method of coating deposition based on the generation of highly ionized metal plasma flows by an arc discharge is widely used to synthesize thin functional coatings \cite{1, 2}. The coatings are formed due to condensation of the plasma of the eroding cathode material on the substrate surface. The cathode material can be in fact any metal, alloy or metal-base composite. With a reactive gas present in the discharge gap, a layer based on the compounds of the cathode material and working gas elements (nitrides, oxides, and carbides) is synthesized. The high degree of ionization of the vacuum-arc plasma and the possibility of controlling the coating synthesis parameters (working gas pressure, discharge current, bias voltage, etc.) over wide limits make it possible to produce a desired effect on the structural, physical, and mechanical characteristics of the condensate. A promising application of the vacuum-arc coating deposition technique is the production of coatings with crystallites less than 100~nm in size. Coatings of this type feature superhardness ($\geq$40~GPa), high resistance to wear and oxidation, etc. Experiments on the plasma-assisted vacuum-arc coating deposition have shown that the grain size can be decreased by adding small amounts of doping elements (Cu, C, Si, Al), which restrict the grain growth in the coating base material during the coalescence of nucleation centers, to the coating composition \cite{3, 4}. Results of the pioneering studies on this subject carried out with the use of CVD processes \cite{5} and magnetron sputtering \cite{6} were published in the late 90-s. Now extensive investigations on the production of multicomponent coatings, such as Ti-Si-N, Ti-Al-N, Zr-Cu-N, and Ti-Si-Al-N, are under way \cite{7,8,9,10}.

There are some methods of the generation of multicomponent arc-discharge plasma for the formation of nanostructured nitride coatings, such as:

(1) production of several metal plasma flows (several single-element cathodes) in the presence of nitrogen reactive gas and

(2) evaporation of composite cathodes whose material includes several elements (mosaic cathodes, sintered powder cathodes, cathodes produced by self-propagating high-temperature synthesis).

Each of these methods has some advantages and disadvantages and, undoubtedly, demands optimization of the parameters of the coating deposition process. In our previous experiments \cite{11,12,13} we have revealed that Ti-Cu, Ti-Si, and Ti-Al composite cathodes produced by sintering metal powders in vacuum \cite{14} are best suited for the synthesis of superhard nitride coatings by the vacuum-arc deposition technique.

Alongside with extensive studies on the plasma-assisted synthesis of nanocrystalline coatings, investigations of their structure-phase composition, properties and of the effect of the doping element on the way by which nanostructurization of the coating occurred are carried out. Two basic models of the structure formation in these coatings depending on their phase composition have been proposed. A detailed description of the model classification is given elsewhere \cite{15}. In the first case, a coating of the nc-MeN/hard phase type is formed with a-Si$_{3}$N$_{4}$, BN, etc. playing the part of the hard phase. For instance, in Ti-Si-N coatings, nanosized crystallites of the basic phase (TiN) are embeded in an amorphous silicon nitride (Si$_{3}$N$_{4}$) matrix. In the second case, when metals which do not form compounds with nitrogen, such as Cu, Ni, Y, Au, etc., are added to MeN coatings, atoms of the doped element surround crystallites of the basic element nitrides, thus restricting their growth on the nanometer scale, and a coating of the nc-MeN/soft phase system (Me = Ti, Zr, Cr, Ta, etc.) is formed. Unfortunately, the capabilities of the existing research techniques restrict the scope of investigation of nanostructured and amorphous materials. Therefore, direct verification of the proposed models of nanostructurization of multicomponent coatings with a certain coating is not always possible, and the use of complex procedures, such as those based on the properties of synchrotron radiation, is required.

This paper presents a study on the structure, phase composition, and element composition of nitride coatings synthesized on metal substrates by evaporating Ti cathode and Ti-Cu composite cathode. The goal of the study was to elucidate the effect of the doping elements on the features of the formation of nanocrystalline structure, on the basic phase crystallite size, and on the properties of the coatings.

\section{The test material and the experimental procedure}

The deposition of nitride coatings in low-pressure arc discharges was carried out on a plasma-ion set-up equipped with a standard arc evaporator and PINK, an original gas-plasma generator developed at the Institute of High Current Electronics, SB RAS (Tomsk) \cite{16}. Additional ionization of the working gas by means of a plasma source with filament cathode made it possible to increase the efficiency of preliminary cleaning of the specimen surface and to realize the formation of nitride compounds under the conditions of plasma assistance.

The generation of multicomponent plasma and the subsequent condensation of coatings were carried out by evaporating cathodes of the compositions Ti containing 12~at.~\% Cu in a nitrogen medium. To compare the coatings obtained and the widely used two-component coatings by their structure-phase and element composition and by mechanical properties, conventional TiN coatings were produced and investigated. Beryllium foil of thickness 0.5~mm, WC-8\%Co hard alloy, and AISI 304 steel were used as substrates for X-ray analysis, mechanical, and TEM investigations, correspondingly. After mechanical grinding, polishing, and washing in an ultrasonic bath, the specimens were placed, on a substrate holder, in a vacuum chamber at a distance of 300~mm from the cathode. Immediately before the deposition of coatings, the surface of specimens was cleaned and activated by bombardment with accelerated argon ions at a negative substrate potential of $\approx$1~kV. During the bombardment, the specimen surface heated up to ~300$^{\circ}$C.

To improve the adhesive characteristics of a nitride coating, its formation was preceded by deposition of a sublayer of thickness about 100 nm by evaporating the cathode material in an argon medium. Synthesis of all multicomponent coatings was carried out in the following parameter ranges: $U_{b} = - (100-300)$~V, $p = 0.3-0.4$~Pa, $I_{d} = 50-100$~A, and $T = 300-400^{\circ}$C. The coating thickness was 3-5~$ \mu$m at a coating growth rate of 1-3~$\mu$m/h.

Investigations of the deposited coatings were carried out by the following methods: optical microscopy (OLYMPUS GX71), transmission (EM-125) and scanning electron microscopy (Philips SEM 515 equipped with EDAX ECON IV, an element composition microanalyzer), micro- and nanoindentation (PMT-3, NHT-S-AX-000X Nano Hardness Tester), scratch testing (MST-S-AX-000 Micro-Scratch Tester), and the Calotest method. The phase and element composition was investigated on powder diffraction stations, on the synchrotron radiation (SR) channels of the VEPP-3 (RFA-SR, PRD SR) energy storage ring and on the PDIFF beamline of the ANKA SR source.

\section{Results and discussion}

\begin{figure}\
 \centerline{\includegraphics [width=8.5cm]{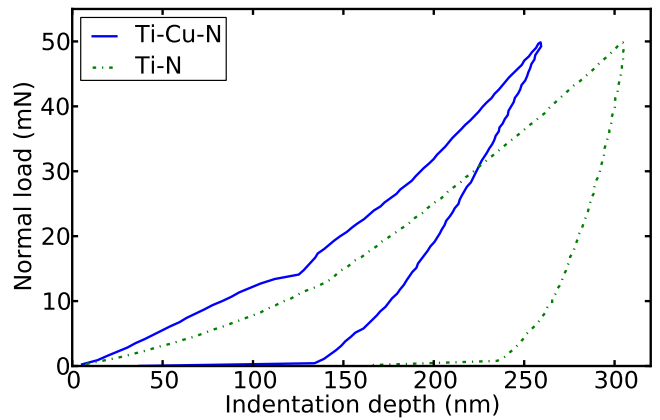}}
 \caption{Loading-unloading curves for coatings produced by evaporating Ti and Ti-12 at.~\% Cu cathodes in arc-discharge plasma.}
 \label{1}
\end{figure}

Measurements of the micro- and nanohardness of the test coatings were performed in order to investigate the influence of crystallite size on mechanical properties of the samples. The measurements performed at a normal load of 500 and 50~mN, respectively, have shown that the hardness of the coatings produced by evaporating composite cathodes is greater than the hardness of titanium nitride ($\approx$25~GPa) by a factor of $\approx$1.5-2. The coatings obtained with the use of Ti-12~at.~\%~Cu cathodes possess superhardness ($\approx$40~GPa). Analysis of the loading-unloading curves (Fig.~\ref{1}) obtained by the nanoindentation method has given the elastic strain of the test coatings. The greatest residual strain of 75\% was observed for a titanium nitride coating and the least of 50\% for Ti-Cu-N coatings. The degree of elastic recovery of the surface shape for the coatings formed by evaporating powder cathodes was 2 times greater than that for the TiN coatings. The Young modulus of the multicomponent coatings was in the range 500-550~GPa.

The scratch-test method was used to determine the critical load at which destruction of a coating begins. For TiN, this quantity was about 3.6~N, whereas the destruction of Ti-Cu-N  coatings of 6.0~N begins at a critical load 2 times greater than that for TiN coatings.

\begin{figure}\
 \centerline{\includegraphics [width=8.5cm]{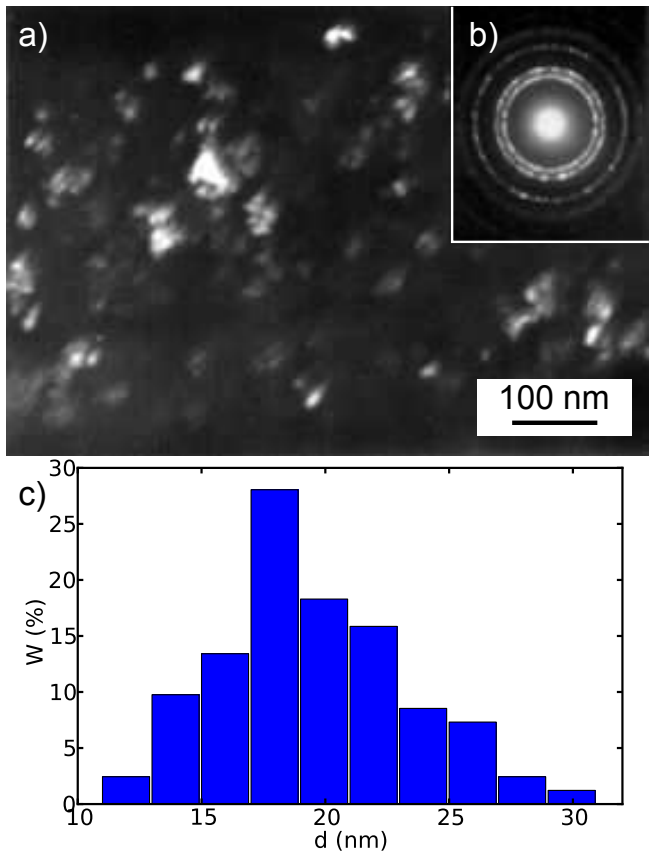}}
 \caption{(Top) Image of the structure of a Ti-Cu-N coating formed by evaporating Ti-12 at.~\% Cu composite cathode in a nitrogen: dark field in the reflection of a type 111 ring of TiN (a), electron diffraction pattern (b). (Bottom) TiN crystallite size distribution in the Ti-Cu-N coating obtained by evaporation of Ti-12at.~\% Cu cathode (c).}
 \label{2}
\end{figure}

With the use of transmission electron diffraction microscopy of thin foils it has been found that the coatings formed by evaporation of composite cathodes consist of nanosized crystallites oriented randomly relative to each other. This follows from the strongly pronounced ring structure of the electron diffraction patterns [Fig.~\ref{2}(b)]. Electron diffraction analysis has shown that the crystallites of the coatings consist of $\delta$-TiN. The measurements of size of coating crystallites were carried out on dark-light images, the average size of crystallite was estimated by the methods of statistical analysis \cite{17}. The size of crystallites in the coatings containing copper is equal to 10-30~nm [Fig.~\ref{2}(a), Fig.~\ref{2}(c)]; the coatings produced by evaporating pure titanium have crystallites of average size 100~nm.

The investigations performed by the above methods gave insufficient information to construct a model of the processes responsible for the variations of the coating properties and to answer the major questions of which is the role of the impurity atoms in the formation of the nanocrystalline structure of synthesized coatings, where they are localized, and whether they form their own crystallographic phase.

It is necessary to notice, that earlier the composite cathode material was investigated \cite{14}. It was showed that copper in cathode material is situated uniformly mainly on boundary of main phase ($\alpha$-Ti) or in compound with Ti (CuTi$_{2}$). The size of titanium based phases is in the range of 3-20~$ \mu$m for Ti-12at.~\% Cu cathode.  The size of cathode spots of vacuum arc is in range of 50-300~$ \mu$m. This fact is evidence of uniformly evaporation of composite cathode elements. 

\begin{figure}\
 \centerline{\includegraphics [width=8.5cm]{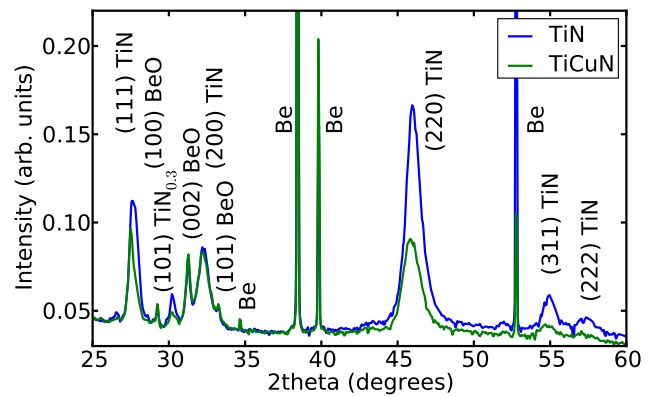}}
 \caption{Diffraction patterns of TiN coatings without doping (TiN) and doped with Cu (Ti-Cu-N) obtained for the photon energy of 10.5~keV}
 \label{4}
\end{figure}

\begin{figure}\
 \centerline{\includegraphics [width=8.5cm]{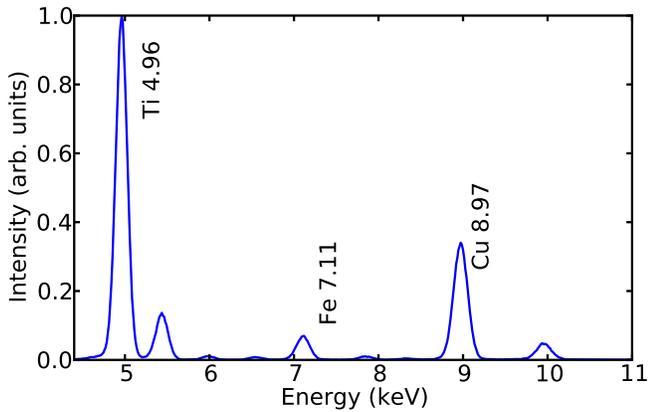}}
 \caption{X-ray fluorescent spectra of Ti-Cu-N specimens near the K-edges of Ti and Cu obtained for the excitation with 20~keV photons.}
 \label{5}
\end{figure}

Investigations of hardening layers based on titanium nitride with additional elements crystalline structure and element composition have been carried out on the stations of the store powder diffraction and x-ray fluorescent element analysis (VEPP-3) \cite{18} and of the PDIFF bamline of ANKA synchrotron radiation source \cite{19}. We rely on the data for the construction of a model of the coatings synthesis. The results of these investigations are given in Fig.~\ref{4} and Fig.~\ref{5}.

Figure~\ref{4} presents diffraction patterns of TiN coatings without admixture and doped with Cu on a beryllium substrate of thickness 0.5~mm. The beryllium substrate shows up in all diffraction patterns in the form of very narrow strongly pronounced reflections. The reflections from the synthesized coating structure are wider owing to their nanocrystalline structure. As well known for powder diffraction \cite{20}, the physical widening of reflections, $\beta$, can be used to estimate the crystallite size in the direction perpendicular to the reflecting plane with indices hkl:

\begin{equation}
\label{e1}
D_{hkl}=n \lambda / \beta \cos \theta
\end{equation}

where $D$ is the size of coherent-scattering region in angstroms, $\lambda$  is the radiation wavelength, $\theta$ is the scattering angle, $\beta$ is the physical line widening in the diffraction pattern in radians (on the 2$\theta$ scale), and n is a coefficient depending on the particle shape and close to unity.

Calculations of crystallite sizes using Eq.~\ref{e1}, where $\beta$ was computed by approximating the reflection lines by the Gauss method, are in agreement with the crystallite sizes measured by electron microscopy methods.

The diffraction patterns of the layers of titanium nitride and of titanium nitride doped with copper are identical [see Fig.~\ref{4}]. The location of reflections testifies to the presence of TiN, TiN$_{0.3}$, and Ti phases in the coatings. Shifts of reflections are not observed in the diffraction patterns obtained for specimens doped with copper. These data allow the conclusion that copper atoms do not form compounds with titanium or nitrogen and, equally, do not form their own crystalline phase. As the data of x-ray fluorescent analysis presented in Fig.~\ref{5} confirm the presence of copper in the test coating in amounts corresponding to its content in the composite cathode evaporated in an arc discharge (Ti-12~at.~\% Cu), it can be concluded that copper is in the amorphous state at the crystallite boundaries. No signal from heavy elements is present in fluorescence data. An stand alone peak at 85~keV is attributed to Be windows. Therefore, we conclude that the only time it takes for copper atoms to form a closed sheath around a growing TiN crystallite determines the time of growth of the crystallite and, hence, its size.

\begin{figure}\
 \centerline{\includegraphics [width=8.5cm]{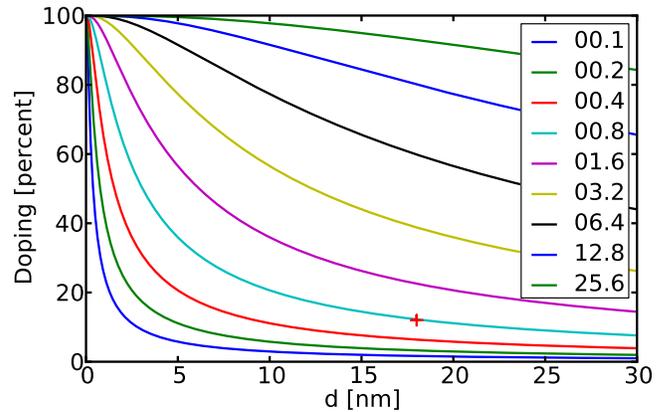}}
 \caption{Simulation of maximum grain size at a given doping concentration. The lines correspond to different coverage needed to prevent further crystallite growth in nm. Our experimental conditions are indicated by a cross.}
 \label{theory}
\end{figure}

In order to explain the observed phenomenon we developed an intuitive model of the growth process. The model assumes that a layer of finite thickness should be formed around a growing grain in order to prevent further growth of the grain. Spherical form of the grain is chosen as a statistical approximation. For a given grain radius $R$ the volume of the grain is calculated as $V=\frac{4}{3} \pi R^{3}$ similarly the volume of a cover layer with thickness $dR$ is calculated as $dV=\frac{4}{3} \pi ((R+dR)^{3}-R^{3})$. The concentration of doping material can be deduced as:

\begin{equation}
\label{model}
x=1-\frac{R^{3}}{(R+dR)^{3}}
\end{equation}

Plot of the grain size dependance on the doping material concentration required to completely cover a growing grain for different cover layer thicknesses shown in Fig.~\ref{theory}. From the figure it is easy to see that the grain size will grow with decrease in the dopant concentration until other processes will not start to dominate limiting the size of the grain. In the case of grain sizes below 5~nm the dependance might be slightly modified by triangulation correction.

If the grain sizes and doping concentrations are known the cover layer thickness is expressed as:

\begin{equation}
\label{model1}
dR=R (\frac{1}{\sqrt[3]{1-x}}-1)
\end{equation}

For our experimental conditions, with 18~nm average grain size and  12~at.~\% Cu concentration we find 0.74~nm cover layer thickness (cross in the Fig.~\ref{theory}) corresponding to 2.8 single atomic layers coverage of the grain (Atomic radius of copper is equal to 132~pm).

\section{Conclusions}

By method of transmission electron diffraction microscopy of thin foils the formation of the nanocrystalline coatings produced by evaporating Ti-Cu composite cathodes in low-pressure arc discharges in the presence of nitrogen plasma was revealed. The coating crystallites consists of $\delta$-TiN. For the copper-containing coatings the average crystallite size is 10-30~nm; the coatings produced by evaporating pure titanium have crystallites of average size 100~nm.

With the powder diffraction method using synchrotron radiation it has been found that copper does not participate in the formation of crystal phases in the coating; that is, they are in the amorphous state. It has been revealed that the copper concentration in the Ti-Cu-N coatings synthesized by the vacuum-arc method coincides with that in the evaporated cathode (12~at.~\%).

The data obtained confirm the model of nanostructurization of coatings based on titanium nitride, according to which the nanocrystallization of these materials occurs as an amorphous sheath of doping elements (copper) is formed around of TiN crystallites, preventing their further growth.

\section*{acknowledgments}
The work was partly supported by SB RAS under Integration Project No. 43, by RFBR under Grant No. 08-08-92207-NSFC-a, and by Presidium RAS under Project No. PP27/09.

\end{document}